# Effect of symmetry on the electronic DOS, charge fluctuations and electron-phonon coupling in carbon chains

[1]C.H.Wong, [2]J.Y.Dai, M.B.Guseva[3], V.N. Rychkov[1], [1]E.A.Buntov, [1]A.F.Zatsepin

[1]Institute of Physics and Technology, Ural Federal University, Ekaterinburg, Russia

[2]Department of Applied Physics, The Hong Kong Polytechnic University, Kowloon, Hong Kong (China)

[3]Faculty of Physics, Moscow State University, Moscow, Russia.

**Abstract:**

A theoretical model is provided to address the parameters influencing the electronic properties of kink-structured carbon chain at 0K. It is studied by the principle of DFT and solving the numerical 1D time-independent Schrödinger equation of electron and phonon simultaneously. Two different lengths of branches $A$ and $B$, are occupied alternatively to generate the asymmetric carbon chain. The ratio of the asymmetric branch length, $R_{AB} = A/B$, plays an important role in the electronic density of states $DOS$ around Fermi level $E_F$. The highest $DOS(E_F)$ occurs if the $R_{AB}$ equals to 2 and while the Fermi level coincides with the Von-Hove singularity at $R_{AB} = 3$. The location of the singularity point relative to the $E_F$ is controllable via branch length interestingly. By comparison with the symmetric case, tuning the branch length asymmetrically shows a stronger impact to shift the $E_F$ to the singularity point.

The numerical solution of the 1D time independent Schrödinger equation of phonon indicates that the kink reinforces the charge fluctuations but the fluctuations are minimized when the $R_{AB}$ goes up. Based on the simulation results, the electron phonon coupling of the carbon nanowire decreases with chain length. In comparison to the symmetric structure, the electron phonon coupling of the asymmetric carbon chain is higher. The maximum electron phonon coupling of the asymmetric carbon chain takes place at $R_{AB} = 2$. However, the weakening of the electron phonon coupling is observed owing to ultrahigh concentration of kinks. The reduction of the electron phonon coupling of the carbon chain is occurred under pressure.

**Introduction:**

The linear carbon chain like carbyne may give impressive mechanical strength [1-5]. Manufacturing long carbon chain is very challenging. However the 6400 atom carbon chain [2] with the incredible elastic modulus of about 53TPa, was successfully measured despite the chain still needs to be stabilized by double wall carbon nanotube (DWCNT). Therefore it is valuable to investigate the physical properties of the carbon nanowire at finite chain length [6]. The elastic modulus of the short linear carbyne containing from 2 to 21 atoms was studied by *ab-initio* methods and demonstrated that the largest elastic modulus came from 5 carbon chain in a supercell [5]. If the elastic modulus of the carbon chain varies with the length of nanowire, the electron-phonon coupling may follow because the elastic modulus depends on the bonding of materials and consequently influences the lattice vibrations [7]. Based on the arguments above, our group intends to investigate the electron phonon coupling as a function of chain lengths and pressures.

The occurrence of the stronger electron phonon coupling is observed in the curvature-induced superconductivity [9]. Kink is one of the indispensable ingredients to create curvature of the carbon chain locally which may be able to strengthen the electron phonon coupling. The exact locations of the kinks are hardly to be controlled and therefore the branch lengths may not unique along the nanowire. What happen if the lengths of branches are not symmetric? Is it harmful to the electron phonon coupling? It inspires us to examine the electron phonon coupling of the asymmetric kink structured carbon chain. For instance, Figure 1 shows that the carbon-4 contains two types of branches. Under this circumstance, the ratio of branch length, $R_{AB} = \frac{A}{B} = \frac{536\,pm}{134\,pm}$ equals to 4 where $A$ and $B$ refer to the corresponding branch length. The asymmetric ratio of $R_{AB} \neq 1$ is possible to perturb the lattice vibration nearby the kinks that may be able to intensify the electron-phonon interaction. As a result, it motivates us to investigate the electron phonon scattering matrix between the symmetric and asymmetric carbon chain. The study of the $DOS$, charge fluctuations and electron phonon coupling of the asymmetric carbon chain will be compared in various $R_{AB}$. Strengthening of the electron phonon coupling via tuning the Fermi-level to the Von-Hove singularity of the $DOS$ is always optimistic [10]. Therefore we are going to create a method to shift the Fermi level to the singularity without applying pressure or gate voltage on the carbon chain. As the kink density also affects the lattice vibrations, we will model the charge fluctuations as a function of kink angles. In this article, all electron phonon coupling strengths and charge fluctuations are relative to the straight carbon chain (kink-less) with the periodic bond length of 134pm.

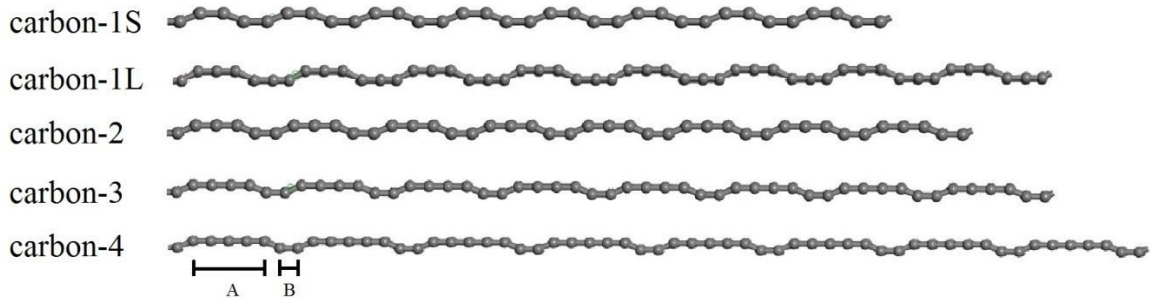

Figure 1: All carbon are connected by double bond length of 134pm and the kink angle is 30 degree (unless otherwise specified). For clarity reason, only 8 basis are shown here. The symmetric branch length of 134pm and 268pm are named as carbon-1S and carbon-1L respectively with $R_{AB} = 1$. The asymmetric branch ratio, $R_{AB}$ of 2, 3 and 4, refers to carbon-2, carbon-3 and carbon-4 are respectively.

**Theory:**

Assume the i$^{th}$ carbon is located at $R_i$ relative to the equilibrium position $R_i^0$, the electron wavefunction $\psi(r)$ is influenced by the Coulomb potential $V(r-R_i)$ and therefore the electron - lattice interaction is given by [9,11]

$$H_{e-ph} = \sum_i \int \psi^\dagger(r) V(r-R_i) \psi(r) dr. \qquad (1)$$

In the light element such as carbon with Z =6, we ignore the spin-orbital effect [7]. As the vibrational amplitude of lattice $u_i$ is rather weak at low temperatures, the first order approximation is applicable. We rewrite the Hamiltonian as

$$H_{e-ph} \sim \sum_i \int \psi^\dagger(r) V(r-R_i^0) \psi(r) dr + \sum_i \int \psi^\dagger(r) [u_i \cdot \nabla V(r-R_i^0)] \psi(r) dr \qquad (2)$$

However, the only important term we focus on is $H_{int} = \sum_i \int \psi^\dagger(r) [u_i \cdot \nabla V(r-R_i^0)] \psi(r) dr \qquad (3)$

The atomic displacement $u_i \propto \frac{1}{\sqrt{Nm}} \sum_q U_q e^{iqR\ell}$ where $U_q \propto \frac{1}{\sqrt{2\omega}} (a_q + a^\dagger_{-q})$ is related to the atomic mass $m$, wavevector of phonon $q$, vibrational frequency $\omega$, total number of atom $N$ and integer $\ell$. The $U_q$ is defined in the creation $a_q$ and annihilation $a^\dagger_{-q}$ operator [9,11]. Based on Bloch Theorem and the property of linear combination of the wavefunction [8], the $\psi(r)$ is expressed as

$\psi(r) = \sum c_k \phi_k(r)$ and $\phi_k(r) = e^{-ik \cdot R_i^0} \phi_k(r + R_i^0)$ where $c_k = <\phi_k^*|\psi>$

Rearranging the terms gives $H_{int} = \sum_{kk'q} c_{k'}^\dagger c_k (a^\dagger_{-q} + a_q) \int \phi_{k'}^*(r) \nabla V(r-R_i^0) \phi_k(r) dr \qquad (4)$

Charge density fluctuation is always created to influence the electrons whenever the phonon is excited [11]. To encounter the screening effect, it is necessary to multiply the reciprocal of permittivity $\frac{1}{\varepsilon}$ in order to obtain the corrected electron phonon interactions.

$$H_{int}^{corrected} = \sqrt{\frac{N}{2m\omega\varepsilon}} \sum_{kk'q} c_{k'}^\dagger c_k (a^\dagger_{-q} + a_q) \int \phi_{k'}^*(r) \nabla V(r-R_i^0) \phi_k(r) dr \qquad (5)$$

We solve the 1D time independent Schrödinger equation in combination with the electronic band diagram to compute the electron wavefunction numerically and therefore $\int \phi_{k'}^*(r) \nabla V(r-R_i^0) \phi_k(r) dr$ is known. The energy band diagram and density of state of electron of the infinite long carbon chain are determined by the Dmol$^3$ package in Material-Studio 7 in which the

Harris functional and Local Density Approximation (LDA) are taken into account [13,14,15,16] in order to reduce computational cost. Assume the electronic band diagram does not change with chain length. The inter-chain separation is 1340pm so that we are likely modelling the isolated 1D carbon chain. We will compare the simulation results based on the GGA functional in CASTEP at the end of the article.

As the effective atomic number of carbon in nanomaterials can be changed due to curvature [11, 16], this phenomenon should also occur in the kink-structured carbon nanowire. The increase of the attractive potential in the Schrödinger equation of the carbon chain in the presence of kinks is estimated with help of the mean electrostatic energy per basis [9,17]. When the electron moves inside the basis, the sum of the electrostatic energy $U_{total}$ between the oppositely charged particles is calculated. As the location of atoms are different between the linear and kink-structured carbon chain, the $U_{total}$ in these two cases are not the same and presumably varies the effective atomic number $Z$ [9,17]. Figure 1 shows different $R_{AB}$ along the carbon chain. The corrected $Z$ of basis in the carbon-1S and carbon-4 due to kinks are raised from 6 to 6.1008 and 6.0906 respectively. As a result, the kink-structured carbon nanowire will be considered as a system of the linear chain of basis. Periodic boundary condition is applied so that the attractive potential acting on the electrons is repeatable across the basis.

Another 1D time-independent Schrödinger equation is solved in order to obtain the wavefunction of phonon in various vibrational quantum numbers according to the dominant acoustic modes and the simple dispersion relationship in the cumulene phase [18]. As the phonon-phonon scattering is weak in the isolated cumulene chain [18] at low temperatures, we estimate the ratio of the collective vibrational frequencies with and without kinks by resolving the spring constants $k_i$ into transverse and longitudinal direction [9] based on 1D harmonic oscillator. The advantage of our algorithm is that the absolute value of the spring constant is not important because we only focus on the relative electron phonon coupling strength of the carbon chain

$$S_{e-ph} = \frac{H_{int}^{corrected}(kink)}{H_{int}^{corrected}(linear)} \tag{6}$$

In other word, there is a relative change in the spring constant after the kink structure is introduced. The charge fluctuations due to atomic motion, $\Delta n_{atomic} = nZ\nabla \cdot u$, hold the key to determine $\sum c_k^\dagger c_k$ where $n$ is the charge concentration [7,9,11]. Again we only focus on the relative charge fluctuations $C_F$.

$$C_F = \frac{\Delta n_{atomic}(kink)}{\Delta n_{atomic}(linear)} \tag{7}$$

The pressure effect on the carbon chain is studied via monitoring the bond length and hence the simulation of the compressed nanowire is computed by updated Fermi energy, electronic band diagram, corrected $Z$ …etc.

**Results and Discussion:**

Figure 2 compares the $DOS$ of the symmetric carbon chains as a function of electron energies under $R_{AB} = 1$. Increasing the branch length symmetrically from 134pm to 268pm pushes the $DOS(E_F)$ about 2.3 times higher. Their Von-Hove singularity points [19,20] in the $DOS$ almost align at the same electron energy and therefore it is ineffective to move the Fermi-level to the singularity point by tuning the branch length symmetrically. Figure 3 provides a predominant method to close the gap between the Fermi-level and singularity point by adjusting the branch length asymmetrically. In case of $R_{AB} \neq 1$, the $E_F$ of the carbon chains are much closer to the Von Hove singularity points when compared to the symmetric case. The $DOS(E_F)$ of $R_{AB} = 3$ coincides with the singularity point which is an appreciative phenomenon to enhance the electron phonon coupling. The peaks in $DOS$ also move similarly in the GGA functional of the CASTEP. However, the most effective way to strengthen the electron phonon coupling is likely to push the Fermi level by 0.06eV in the carbon-2. In general, the $DOS(E_F)$ of the symmetric carbon chain is the lower than the asymmetric nanowire as shown in Figure 2 and Table 1. Table 2 reveals that the $C_F$ reduces with increasing $R_{AB}$. It is credited to the weaker perturbation of phonon due the lowering of kink density.

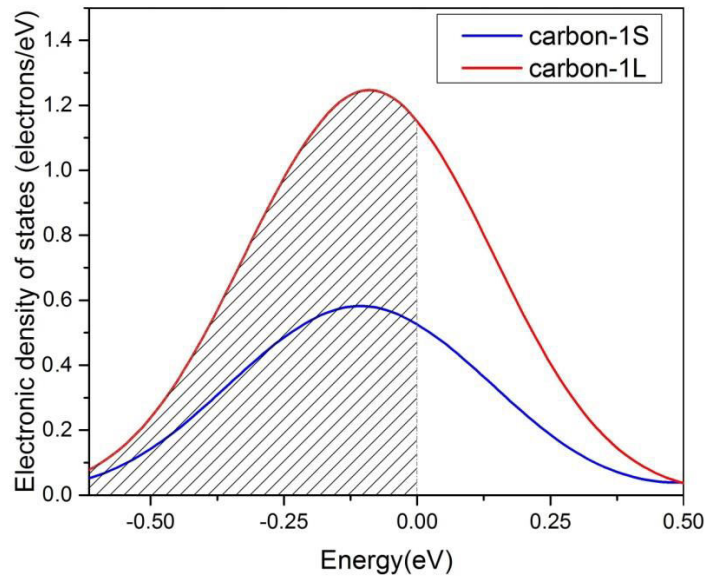

Figure 2: The electronic density of states $DOS$ in the symmetric carbon chains. The Fermi-level is shifted to 0 eV for convenience. The shaded region refers to the filled states

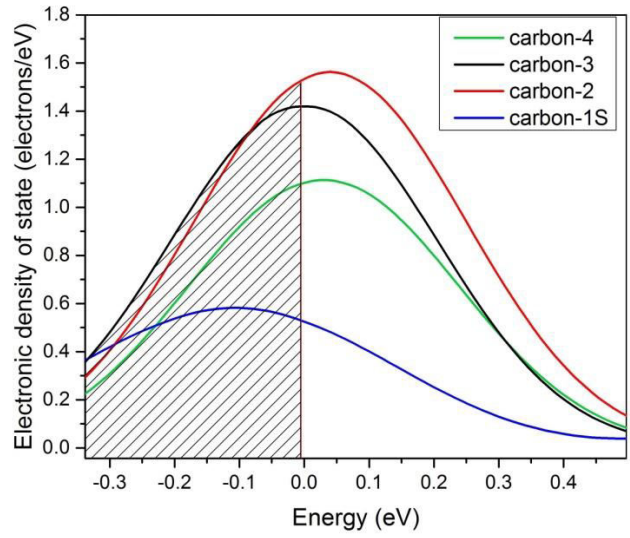

Figure 3: The electronic density of state $DOS$ of the carbon chains in various $R_{AB}$. The Fermi-level is adjusted to 0 eV for better readability.

Table 1: Compare the ratio of the $DOS$ at Fermi level relative to the linear carbon chain

|  | $R_{AB}$ (relative to 134pm) | $DOS(E_F)$ |
|---|---|---|
| carbon-1S | 1 | 0.52 |
| carbon-2 | 2 | 1.53 |
| carbon-3 | 3 | 1.42 |
| carbon-4 | 4 | 1.09 |

Table 2: Compare the relative charge fluctuations $C_F$ in various $R_{AB}$

|  | $R_{AB}$ (relative to 134pm) | $C_F$ |
|---|---|---|
| carbon-1S | 1 | 1.19 |
| carbon-2 | 2 | 1.14 |
| carbon-3 | 3 | 1.11 |
| carbon-4 | 4 | 1.09 |

Figure 4 illustrates that the $C_F$ becomes stronger when the kink angle is bigger. In the case of isolated cumluene, the longitudinal acoustic mode is dominant [18]. However, the tiny amount of the orthogonal vibration nearby the kinks is likely disturbing to the longitudinal mode and presumably the charge fluctuation goes up. In the extreme case of $R_{AB} \gg 1$, the geometry should returns to the linear kink-less carbon chain and the $C_F$ is expected to drop to 1. Figure 5 demonstrates that the $S_{e-ph}$ of the carbon-4 increases with the length of nanowire but the $S_{e-ph}$ show no sign of saturation up to 11.5nm. The length of nanowire is simulated only up to about 12nm because of minimizing the computational cost. If the 1D nanowire is long enough, the $S_{e-ph}$ should be length-independent [10]. However the length of 11.5nm is still comparable to the margin between 0D quantum dot and 1D nanowire [22,23] and therefore the saturation of the $S_{e-ph}$ of the short carbon chain is not noticeable in Figure 5.

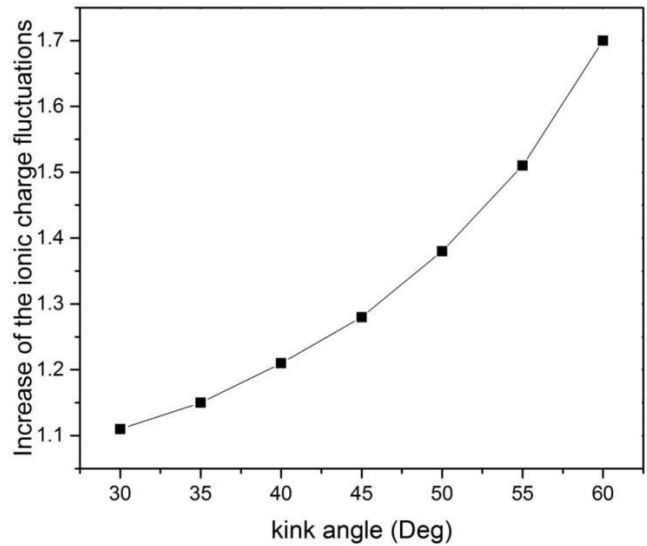

Figure 4: The $C_F$ as a function of kink angles in the carbon-3. The kink angle is relative to the linear carbon chain where the pivot angle between the three consecutive carbons is 0 degree.

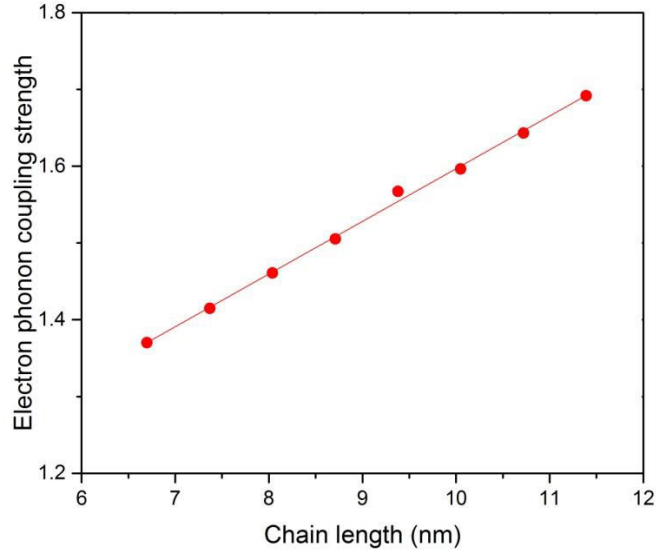

Figure 5: The electron phonon coupling $S_{e-ph}$ of the carbon-4 as a series of chain lengths

The highest kink density, carbon-1S, shows the weakest electron phonon coupling unexpectedly despite the charge density is largest as shown in Table 3. The $S_{e-ph}$ is less than 1 which implies that the electron phonon coupling of the carbon-1S is surprisingly weaker than the linear kink-less carbon chain. The main reason is the obvious reduction of the $DOS(E_F)$ in the carbon-1S which outweighs the contribution from the stronger electrostatic attraction. On one hand, ultra-long branch will reduce the $S_{e-ph}$ to 1 if the kink density is too low. On the other hand, the ultrahigh concentration of kink also put the reinforcement in the electron phonon scattering in disadvantage based on the reduced $DOS$. Interestingly, the optimal electron phonon interaction of the symmetric kink-structured carbon nanowire may exist somewhere in the moderate kink density. The electrons and phonons inside the asymmetric carbon-2 interact tighter than either the carbon-1S or carbone-1L as illustrated in Table 3. The appearance of the stronger electron phonon interaction in the asymmetric carbon-2 should be originated from the higher $DOS(E_F)$ and the asymmetric distortion of phonon.

Table 3: Electron phonon coupling - symmetry vs asymmetry

|  | Type | Branch length(pm) | Electron phonon coupling $S_{e-ph}$ |
|---|---|---|---|
| Carbon-1S | Symmetric branch | 134 | 0.70 |
| Carbon-1L | Symmetric branch | 268 | 1.31 |
| Carbon-2 | Asymmetric branch | 134 & 268 alternatively | 2.14 |

In the absence of pressure, the strongest $S_{e-ph}$ occurs in the carbon-2 with the bond length of 134pm as shown in Figure 6 and the decrease of the $S_{e-ph}$ is observed if $R_{AB} \geq 3$. The $S_{e-ph}$ is likely closed to 1 if the $R_{AB} \gg 1$. The main argument is based on the tendency of charge fluctuation and $DOS(E_F)$ which should reduce to the linear case for extremely large $R_{AB}$. The $S_{e-ph}$ of the carbon-1S is the weakest. It is again explained by the lowest $DOS(E_F)$ in the carbon-1S. Hence the $DOS(E_F)$ plays a major role to influence the $S_{e-ph}$ of 1D material [10] because the $DOS$ in combination with Fermi-Dirac statistic affect the probability of electron to interact with phonon.

Fermi energy $E_f$ is a function of pressure $P$. Our DFT data shows that the compressed bond length of 131.3pm corresponds to the shift in Fermi energy around +0.22 to +0.28 eV for all types of carbon nanowires in this article. The electron phonon couplings of the compressed carbon chains are dramatically decreased regardless of the $R_{AB}$. The main reason is due to the significant reduction of the $DOS$ at $E_f + \Delta E_f(P)$ eV in Figure 3. However, if the carbon-2 and carbon 4 are subjected by low pressure to move the Fermi-level by +0.05eV only, the electron phonon coupling may be able to be reinforced by compression due to the rise of $DOS(E_F)$. The feasibility of the superconductivity in carbyne is still controversial. If carbyne does show Type-1 superconductivity such as Pb [25,26], the superconducting transition temperature of the carbyne likely reduces with large pressure. Despite the electron phonon coupling is higher in the asymmetric case, examining the stability in the asymmetric kink structure is indispensable. Studying the energy of the symmetric carbon-1L and asymmetric carbon-3 is more wisely because they hold the same basis size. The energy of the carbon-1L is only 0.18% higher than the carbon-3. It shows that the formation of the asymmetric carbon chain is still possible.

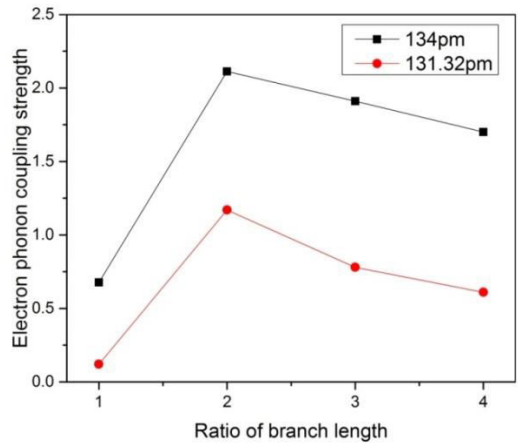

Figure 6: The $S_{e-ph}$ is the strongest in the free standing carbon-2. The significant reduction of the $S_{e-ph}$ are observed in the double bond of 131.3pm. The branch ratio of 1 refers to the carbon-1S.

Any ratio of branch length can be made by a lot of combinations.

For example, $R_{AB} = \frac{268\,pm}{134\,pm} = \frac{2680\,pm}{1340\,pm} = \frac{53600\,pm}{26800\,pm} = 2$. If the branches in the carbon nanowire are extremely long, e.g. $R_{AB} = \frac{268\,mm}{134\,mm} = 2$, the increase of electron phonon coupling due to the asymmetric structure is rare because the size effect of the local lattice vibration in each branch is saturated. As a result, our group keep the branch length B (134pm) as short as possible in order to detect the obvious change in the electron phonon coupling owing to the asymmetric branch structure. It is interesting to boost the electron phonon coupling of the carbon chain. Our proposal is to optimize the length of the symmetric branch which maximizes the electron phonon coupling first. Afterwards, we suggest adjusting the ratio of branch length until we observe the optimal electron phonon scattering of the asymmetric carbon chain. Hopefully the kink-structured carbon chain will be the new group member of superconductivity after optimizing the electron phonon coupling. In spite of having a lot of nanowire systems being modelled by Dmol$^3$ package [29,30], the CASTEP package with much higher computational costs may be more preferable to compute the nanowire system. However, Table 4 draw parallel between the Dmol$^3$ and CASTEP, the simulation results obtaining from these two packages are consistent regardless of the effect of symmetry where the carbon-4 gives the largest $R_{AB}$. In view of this, although some approximations are taken in the calculation, the primary conclusions presented in this article and the observed trends of electron phonon coupling of the asymmetric and symmetric carbon chain should still hold.

Table 4: Data comparison between the Dmol$^3$ and CASTEP.

|  | Type | $S_{e-ph}$ (by Dmol$^3$) | $S_{e-ph}$ (by CASTEP) |
| --- | --- | --- | --- |
| Carbon-1S | Symmetric branch | 0.70 | 0.81 |
| Carbon-1L | Symmetric branch | 1.31 | 1.42 |
| Carbon-4 | Asymmetric branch | 1.71 | 1.69 |

**Conclusion:**

In this article we investigate the physical properties of the carbon chain under symmetric and asymmetric branch. Instead of compressing the sample or injecting gate voltage, a guideline of tuning the Fermi-level to the Von-Hove singularity in the electronic density of state is provided by our DFT modelling. According to our simulation, closing the gap between the Fermi-level and the Von-Hove singularity by the generation of asymmetric kink structure is more wisely than the design of symmetric carbon chain. The Fermi-level of the carbon chain is intrinsically aligned at the Von-Hove singularity if the ratio of the asymmetric branch length is 3 (relative to 134pm).

The charge fluctuates weaker if the branches are too long and hence the electron phonon coupling should move towards 1 in the case of infinitely long branch. The chain-length effect on the electron phonon coupling of the carbon nanowire is studied. The shorter carbon nanowire it has, the weaker electron-phonon scattering it takes. In addition, the electron phonon coupling of the carbon chain can be tuned by the branch ratio and external pressure. We concluded that the asymmetric structure gives a tighter electron phonon coupling of the carbon chain. We hope that these data can help us to understand the origin of the physical properties of the 2D ordered linear chain of carbon and other nanocarbon forms.